\documentclass[12pt]{article}
\usepackage[english,german,french,polish]{babel}
\usepackage[T1]{fontenc}
\usepackage{amsfonts}

\begin{document}

\selectlanguage{english}

\baselineskip 0.75cm
\topmargin -0.6in
\oddsidemargin -0.1in

\let\ni=\noindent

\renewcommand{\thefootnote}{\fnsymbol{footnote}}

\pagestyle {plain}

\setcounter{page}{1}

\pagestyle{empty}

~~~

\begin{flushright}
IFT--04/17
\end{flushright}

{\large\centerline{\bf Can one of three righthanded neutrinos}}

{\large\centerline{\bf be light enough to produce a small LSND effect?{\footnote {Work supported in part by the Polish State Committee for Scientific Research (KBN), grant 2 P03B 129 24 (2003--2004).}}}}

\vspace{0.4cm}

{\centerline {\sc Wojciech Kr\'{o}likowski}}

\vspace{0.3cm}

{\centerline {\it Institute of Theoretical Physics, Warsaw University}}

{\centerline {\it Ho\.{z}a 69,~~PL--00--681 Warszawa, ~Poland}}

\vspace{0.6cm}

{\centerline{\bf Abstract}}

\vspace{0.2cm}

It is shown on the ground of a simple $6\times 6$ neutrino mixing model that one of three conventional 
sterile (righthanded) neutrinos, if light enough, may be consistently used for explaining a {\it small} 
LSND effect. Then, it is still considerably heavier than the three active (lefhanded) neutrinos, so that 
a kind of {\it soft} seesaw mechanism can work. The usual condition that the Majorana lefthanded 
component of the overall $6\times 6$ neutrino mass matrix ought to vanish, {\it implies} the 
smallness of active-neutrino masses versus sterile-neutrino masses, when three mixing angles 
between both sorts of neutrinos are small. In the presented model, the mass spectrum of active neutrinos comes out roughly degenerate, lying in the range $(5\;  -  \;7.5)\times 10^{-2}$ eV, {\it if} there is a {\it small} LSND effect with the amplitude of the order $10^{-3}$ and with the mass-squared splitting $\sim 1\;{\rm eV}^2$. 

\vspace{0.2cm}

\ni PACS numbers: 12.15.Ff , 14.60.Pq , 12.15.Hh .

\vspace{0.6cm}

\ni April 2004 

\vfill\eject

~~~
\pagestyle {plain}

\setcounter{page}{1}

\vspace{0.2cm}

It is well known that the neutrino experiments with solar  $\nu_e$'s [1], atmospheric $\nu_\mu$'s [2], long-baseline accelerator $\nu_\mu$'s [3] and long-baseline reactor $\bar{\nu}_e$'s [4] are very well described by oscillations of three active neutrinos $\nu_e \,,\, \nu_\mu \,,\, \nu_\tau $, where the 
mass-squared splittings of the related neutrino mass states $\nu_1\,,\, \nu_2 \,,\, \nu_3 $ are estimated to be $\Delta m^2_{\rm sol} \equiv \Delta m^2_{21} \sim 7\times 10^{-5}\; {\rm eV}^2$ and $\Delta m^2_{\rm atm} \equiv \Delta m^2_{32} \sim 2.5\times 10^{-3}\; {\rm eV}^2$ [5]. The neutrino mixing matrix $U^{(3)} = \left(U^{(3)}_{\alpha i} \right) \;(\alpha = e, \mu, \tau\;{\rm and}\; i=1, 2, 3)$, responsible for the unitary transformation

\begin{equation}
\nu_\alpha  = \sum_i U^{(3)}_{\alpha i}\, \nu_i \;,
\end{equation}

\ni is experimentally consistent with the global bilarge form


\begin{equation}
U^{(3)} = \left( \begin{array}{ccc} c_{12} & s_{12} & 0 \\ - \frac{1}{\sqrt2} s_{12} & \frac{1}{\sqrt2} c_{12} & \frac{1}{\sqrt2}  \\ \frac{1}{\sqrt2} s_{12} & -\frac{1}{\sqrt2} c_{12} & \frac{1}{\sqrt2}  \end{array} \right) \;,
\end{equation}

\ni where $c_{12} = \cos \theta_{12}$ and $s_{12} = \sin \theta_{12}$ with $\theta_{12} \sim 33^\circ $, while $U^{(3)}_{e3} = s_{13} \exp(-i\delta)$ is neglected according to the negative results of 
neutrino disappearance experiments with short-baseline reactor $\bar{\nu}_e$'s, in particular the  
Chooz experiment [6] that estimates the experimental upper bound for $s^2_{13}$ as $s^2_{13} < 0.03$.

However, the signal of $\bar{\nu}_\mu \rightarrow \bar{\nu}_e $ appearance reported by the LSND experiment with short-baseline accelerator $\bar{\nu}_\mu$'s [7] requires for its interpretation in terms of neutrino oscillations a third neutrino mass-squared splitting, say, $\Delta m^2_{\rm LSND} \sim 1\; {\rm eV}^2$. This cannot be justified by the use of only three neutrinos (unless the CPT invariance of neutrino oscillations is seriously violated, leading to considerable mass splittings of neutrinos and antineutrinos [8]; in the present note the CPT invariance is assumed to hold). The LSND result will be tested soon in the ongoing MiniBooNE experiment [9]. If this test confirms the LSND result, we will need the light sterile neutrinos in addition to three active neutrinos to introduce extra mass splittings.

While the 3+1 neutrino models with one light sterile neutrino are considered to be disfavored by present data [10], the 3+2 or 3+3 neutrino schemes with two or three light sterile neutrinos may {\it a priori} provide a better description of current neutrino oscillations including the LSND effect (for a statistical discussion showing the better compatibility of all short-baseline neutrino experiments within 3+2 models than within 3+1 models {\it cf.} Ref. [11]; in Ref. [12] we argue, however, that the simple 3+2 models are not more effective in this description than the simple 3+1 models: both kinds of them may be consistent with a {\it small} LSND effect having the amplitude of, say, the order $10^{-3}$).

In the present note, we discuss the question to what extent three conventional sterile (righthanded) neutrinos may help to reconcile the possible LSND effect with the well established results of solar and atmospheric oscillation experiments.

To this end, consider the usual neutrino theory, where the Majorana lefthanded component $ M^{(L)}$ of the overall $6\times 6$ neutrino mass matrix $ M^{(6)} = \left( M^{(6)}_{\alpha \beta}\right)\;\;(\alpha, \beta = e , \mu , \tau , e_s , \mu_s , \tau_s)$ is zero:

\begin{equation}
 M^{(6)} = \left( \begin{array}{cc} 0 & M^{(D)} \\ M^{(D)\,T} & M^{(R)} \end{array} \right) \,.
\end{equation}

\ni Here, three active neutrinos $\nu_{e , \mu , \tau} \equiv \nu_{e , \mu , \tau\,L}$ and three conventional sterile antineutrinos $\nu_{e_s , \mu_s , \tau_s} \equiv \left(\nu_{e , \mu , \tau\,R}\right)^c$ form the basis of a 3+3 neutrino model. Then, the overall $6\times 6$ neutrino mixing matrix $ U^{(6)} = \left( U^{(6)}_{\alpha i}\right)$ transforms unitarily flavor neutrinos $\nu_{\alpha}\;\;(\alpha = e , \mu , \tau , e_s , \mu_s , \tau_s)$ into mass neutrinos $\nu_i\;\;(i = 1,2,3,4,5,6)$ :

\begin{equation}
\nu_\alpha  = \sum_i U^{(6)}_{\alpha i}\, \nu_i \;.
\end{equation}

\ni In the flavor representation, where the charged-lepton mass matrix is diagonal, the $6\times 6$ mixing matrix $U^{(6)}$ is at the same time the $6\times 6$ diagonalizing matrix for the $6\times 6$ mass matrix $M^{(6)}$ :

\begin{equation} 
U^{(6)\,\dagger} M^{(6)}{U}^{(6)} = {\rm diag}(m_1\,,\,m_2\,,\,m_3\,,\,m_4\,,\,m_5\,,\,m_6)
\end{equation}

\ni and so, inversely

\begin{equation}
M^{(6)}_{\alpha \beta}  = \sum_i U^{(6)}_{\alpha i}\, m_i\, U^{(6)*}_{\beta i}\;.
\end{equation}

To proceed further we will assume the simple $6\times 6$ neutrino mixing model, where

\begin{equation}
U^{(6)} = \left(\begin{array}{rr} U^{(3)} & 0^{(3)} \\ 0^{(3)} & 1^{(3)} \end{array}\right)
\left(\begin{array}{cc} C^{(3)} & S^{(3)} \\ -S^{(3)} & C^{(3)} \end{array}\right) =  \left(\begin{array}{cc} U^{(3)} C^{(3)} & U^{(3)} S^{(3)} \\ -S^{(3)} & C^{(3)} \end{array}\right) 
\end{equation}

\ni with $U^{(3)}$ given in Eq. (2) and

\begin{equation}
C^{(3)} = \left(\begin{array}{ccc} c_{14} & 0 & 0 \\ 0 & c_{25} & 0 \\ 0 & 0 & c_{36} \end{array}\right) \;,\;
S^{(3)} = \left(\begin{array}{ccc} s_{14} & 0 & 0 \\ 0 & s_{25} & 0 \\ 0 & 0 & s_{36} \end{array}\right) \,,
\end{equation}

\ni where $c_{14} = \cos \theta_{14}$, $s_{14} = \sin \theta_{14}$ and so on. Thus, in Eq. (7)

\begin{eqnarray}
U^{(3)}C^{(3)} & = & \left( \!\!\begin{array}{ccc} c_{12} c_{14} & s_{12}c_{25} & 0 \\ -\frac{1}{\sqrt2} s_{12}c_{14} & \frac{1}{\sqrt2} c_{12}c_{25} & \frac{1}{\sqrt2} c_{36} \\ \frac{1}{\sqrt2} s_{12}c_{14} & -\frac{1}{\sqrt2} c_{12}c_{25}  & \frac{1}{\sqrt2} c_{36} \end{array} \!\right) \,, \nonumber \\
U^{(3)}S^{(3)} & = & \left( \!\!\begin{array}{ccc} c_{12} s_{14} & s_{12}s_{25} & 0 \\ -\frac{1}{\sqrt2} s_{12}s_{14} & \frac{1}{\sqrt2} c_{12}s_{25} & \frac{1}{\sqrt2} s_{36} \\ \frac{1}{\sqrt2} s_{12}s_{14} & -\frac{1}{\sqrt2} c_{12}s_{25} & \frac{1}{\sqrt2} s_{36} \end{array} \!\right) \,.
\end{eqnarray}

\ni Due to Eq. (7) with (9), the unitary mixing transformation $ \nu_i =\sum_\alpha U^{(6)*}_{\alpha i} \nu_\alpha $, inverse to (4), reads explicitly

\begin{eqnarray}
\nu_1 & = & c_{14}\left(c_{12} \nu_e - s_{12} \frac{\nu_\mu - \nu_\tau}{\sqrt2}\right) - s_{14} \nu_{e_s} \,, \nonumber \\ 
\nu_2 & = & c_{25}\left(s_{12} \nu_e + c_{12} \frac{\nu_\mu - \nu_\tau}{\sqrt2}\right) - s_{25}\nu_{\mu_s} \,, \nonumber \\
\nu_3 & = & c_{36} \frac{\nu_\mu + \nu_\tau}{\sqrt2} - s_{36} \nu_{\tau_s} \,, \nonumber \\ 
\nu_4 & = & s_{14}\left(c_{12} \nu_e - s_{12} \frac{\nu_\mu - \nu_\tau}{\sqrt2}\right) + c_{14}\nu_{e_s}\,, \nonumber \\
\nu_5 & = & s_{25}\left(s_{12} \nu_e + c_{12} \frac{\nu_\mu - \nu_\tau}{\sqrt2}\right) + c_{25}\nu_{\mu_s} \,, \nonumber \\
\nu_6 & = & s_{36} \frac{\nu_\mu + \nu_\tau}{\sqrt2} + c_{36}\nu_{\tau_s}\,.
\end{eqnarray}

\ni Here, $\nu_\mu $ and $\nu_\tau$ mix maximally, since $(\nu_\mu - \nu_\tau)/\sqrt2 $ and $(\nu_\mu + \nu_\tau)/\sqrt2 $ do not mix at all. More generally, $\nu_e ,  (\nu_\mu - \nu_\tau)/\sqrt2 , \nu_{e_s} $ and $\nu_{\mu_s}$ do not mix at all with $(\nu_\mu + \nu_\tau)/\sqrt2 $ and $\nu_{\tau_s}$.

Applying Eqs. (6) and (7)  with (9), we obtain

\begin{eqnarray}
M^{(L)} & = & U^{(3)} \left(\begin{array}{ccc} c^2_{14}m_1 + s^2_{14} m_4 & 0 & 0 \\ 0 & c^2_{25} m_2 + s^2_{25} m_5  & 0 \\ 0 & 0 & c^2_{36} m_3 + s^2_{36} m_6  \end{array}\right) U^{(3)\,\dagger}  \,, \\ 
M^{(D)} & = & U^{(3)} \left(\begin{array}{ccc} c_{14} s_{14} (m_4 - m_1) & 0 & 0 \\ 0 & c_{25} s_{25} (m_5 - m_2)  & 0 \\ 0 & 0 & c_{36} s_{36}(m_6 - m_3)  \end{array}\right) 
\end{eqnarray}

\ni and

\begin{equation}
M^{(R)} = \left(\begin{array}{ccc} c^2_{14}m_4 + s^2_{14} m_1 & 0 & 0 \\ 0 & c^2_{25} m_5 + s^2_{25} m_2  & 0 \\ 0 & 0 & c^2_{36} m_6 + s^2_{36} m_3  \end{array}\right) \,.
\end{equation}

\ni Due to Eq. (11), the condition $M^{(L)} = 0$ tells us that 

\begin{equation}
m_1 = - t^2_{14} m_4 \,,\, m_2 = - t^2_{25} m_5 \,,\, m_3 = - t^2_{36} m_6 \,, 
\end{equation}

\ni where $t_{14} = s_{14}/c_{14} = \tan \theta_{14}$ and so on. Then, Eqs. (12) and (13) take the forms 

\begin{equation}
M^{(D)} = U^{(3)}\left(\begin{array}{ccc} t_{14} m_4  & 0 & 0 \\ 0 & t_{25} m_5  & 0 \\ 0 & 0 & t_{36}m_6  \end{array}\right) 
\end{equation}

\ni and

\begin{equation}
M^{(R)} = \left(\begin{array}{ccc} (1-t^2_{14})m_4 & 0 & 0 \\ 0 & (1 - t^2_{25}) m_5 & 0 \\ 0 & 0 & (1 - t^2_{36}) m_6 \end{array}\right) \,.
\end{equation}

\ni Hence, we calculate

\begin{equation}
-M^{(D)} \frac{1}{M^{(R)}} M^{(D)\,T} = U^{(3)} \left(\begin{array}{ccc} m_1/(1-t^2_{14}) & 0 & 0 \\ 0 & m_2/(1 - t^2_{25}) & 0 \\ 0 & 0 & m_3/(1 - t^2_{36}) \end{array}\right) U^{(3)\,\dagger} \,.
\end{equation}

\ni If $t^2_{14} = |m_1/m_4| \ll 1$, $t^2_{25} = |m_2/m_5| \ll 1$ and $t^2_{36} = |m_3/m_6| \ll 1$, as it is the case in the seesaw mechanism, two expressions

\begin{equation}
-M^{(D)} \frac{1}{M^{(R)}} M^{(D)\,T}  \simeq U^{(3)} \left(\begin{array}{ccc} m_1 & 0 & 0 \\ 0 & m_2 & 0 \\ 0 & 0 & m_3 \end{array}\right) U^{(3)\,\dagger} 
\end{equation}

\ni and

\begin{equation}
M^{(R)}  \simeq \left(\begin{array}{ccc} m_4 & 0 & 0 \\ 0 & m_5 & 0 \\ 0 & 0 & m_6 \end{array}\right) 
\end{equation}

\ni describe approximately the Majorana mass matrices for active (lefthanded) and sterile (righthanded) neutrinos, respectively (the second mass matrix is here diagonal). But, {\it a priori},  it is not necessary for the small ratios $t^2_{14} = |m_1/m_4| \ll 1$, $t^2_{25} = |m_2/m_5| \ll 1$ and $t^2_{36} = |m_3/m_6| \ll 1$ to be so drastically small as in the case of seesaw mechanism. We will see that this alternative scenario may be consistently realized, when one of three sterile (righthanded) neutrinos produces a small LSND effect.

In the case of $6\times 6$ mixing matrix $U^{(6)}$ given in Eqs. (7) and (9), we obtain the following neutrino oscillation probabilities in the vacuum, if $x_{31} \simeq x_{32}$, $x_{41} \simeq x_{42} \simeq x_{43}$, $x_{51} \simeq x_{52} \simeq x_{53}$, $x_{61} \simeq x_{62} \simeq x_{63}$ and $c^2_{14} \gg s^2_{14}$, $c^2_{25} \gg s^2_{25}$, $c^2_{36} \gg s^2_{36}$ :

\begin{eqnarray}
P(\nu_e \rightarrow \nu_e) & \simeq & 1 - 4c^2_{12}s^2_{12}\sin^2 x _{21} - 4c^2_{12}s^2_{14}  \sin^2 x _{41} - 4s^2_{12} s^2_{25} \sin^2 x_{51} \,, \\
P(\nu_\mu \rightarrow \nu_\mu) & \simeq & 1  - c^2_{12}s^2_{12} \sin^2 x_{21}  - \sin^2 x_{31} \nonumber \\ 
& & \:\:\:-\,  2s^2_{12}s^2_{14} \sin^2 x _{41}  - 2c^2_{12}s^2_{25} \sin^2 x _{51}  - 2s^2_{36} \sin^2 x _{61} 
\end{eqnarray}

\ni and

\begin{equation}
P(\bar{\nu}_\mu \rightarrow \bar{\nu}_e)  \simeq  2c^2_{12}s_{12}^2 \sin^2 x _{21} + 2c^2_{12} s^2_{12} (s^2_{14} - s^2_{25})(s^2_{14}\sin^2 x _{41} - s^2_{25}\sin^2 x_{51})  \,, 
\end{equation}

\ni where $x_{j i} \equiv 1.27 \Delta m^2_{j i} L/E$ and $\Delta m^2_{j i} \equiv m^2_j - m^2_i$. In Eqs. (20) and (21), quadratic terms with respect to the small parameters $s^2_{14} \,,\, s^2_{25}$ and $s^2_{36}$ are neglected. 

Hence, for solar $\nu_e$'s, Chooz reactor $\bar{\nu}_e$'s, atmospheric $\nu_\mu$'s and LSND accelerator $\bar{\nu}_\mu$'s, where $(x_{21})_{\rm sol} \sim O(\pi /2)$, $(x_{31})_{\rm Chooz} \simeq (x_{31})_{\rm atm} \sim O(\pi /2)$ and $(x_{41})_{\rm LSND} \sim O(\pi /2)$, respectively, we deduce the following oscillation probabilities :

\begin{eqnarray}
P(\nu_e \rightarrow \nu_e)_{\rm sol}\:\: & \simeq & 1 - 4c^2_{12}s^2_{12} \sin^2 (x _{21})_{\rm sol} - 2(c^2_{12}s^2_{14} + s^2_{12}s^2_{25}) \,, \\ 
P(\bar{\nu}_e \rightarrow \bar{\nu}_e)_{\rm Chooz} & \simeq & 1 - 2(c^2_{12}s^2_{14} + s^2_{12}s^2_{25}) \,,  \\
P(\nu_\mu \rightarrow \nu_\mu)_{\rm atm} & \simeq & 1 -  \sin^2 (x_{31})_{\rm atm} -  (s^2_{12}s^2_{14} + c^2_{12}s^2_{25} + s^2_{36}) 
\end{eqnarray}

\ni and

\begin{equation}
P(\bar{\nu}_\mu \rightarrow \bar{\nu}_e)_{\rm LSND}  \simeq  2c^2_{12} s^2_{12} \left(s^2_{14} - s^2_{25}\right) \left(s^2_{14} \sin^2 (x_{41})_{\rm LSND} - \frac{1}{2} s^2_{25}\right) \,, 
\end{equation}

\ni if $x_{21} \ll x_{31} \ll x_{41} , x_{51}, x_{61}$ {\it i.e.},  $m^2_1, m^2_2, m^2_3 \ll m^2_4, m^2_5, m^2_6$. In Eq. (26), it is assumed in addition that $x_{41} \ll x_{51}$ {\it i.e}, $m^2_4 \ll m^2_5$. Of course, for solar $\nu_e$'s the MSW matter mechanism is significant, leading to the experimentally accepted LMA solar solution.

If there is a small LSND effect with the amplitude of the order $10^{-3}$, then due to Eq. (26) we can write

\begin{equation}
\left(s^2_{14} - s^2_{25}\right)^{\!\!1/2} \left(s^2_{14} - \frac{s^2_{25}}{2\sin^2(x_{41})_{\rm LSND}}  \right)^{\!\!1/2} \sim \left(\frac{10^{-3}}{2c^2_{12}s^2_{12}} \right)^{\!\!1/2} \sim 0.049 \,, 
\end{equation}

\ni where $\theta_{12} \sim 33^\circ $ giving $c^2_{12} \sim 0.70$ and $s^2_{12} \sim 0.30$. In the case of $1 \gg s^2_{14} \gg s^2_{25} \gg s^2_{36}$ {\it i.e.}, $ 1 \gg t^2_{14} = |m_1/m_4| \gg t^2_{25} = |m_2/m_5| \gg t^2_{36} = |m_3/m_6|$ (even if $m^2_1 < m^2_2 <m^2_3$), Eq. (27) gives

\begin{equation} 
s^2_{14} \sim \left(\frac{10^{-3}}{2c^2_{12}s^2_{12}}\right)^{1/2} \sim 0.049\,.
\end{equation}

\ni Hence, $ |m_1/m_4|  =  t^2_{14} \sim 0.052 \ll 1$, though this ratio is not so dramatically small as in the seesaw mechanism. If $\Delta m^2_{41} \sim 1\;{\rm eV}^2$, then $|m_4| \sim 1$ eV and we predict that $|m_1| \sim 5.2\times 10^{-2}$ eV. In this case, from the experimental estimates $\Delta m^2_{21} \sim 7\times 10^{-5}\;{\rm eV}^2$ and $\Delta m^2_{32} \sim 2.5\times 10^{-3}\;{\rm eV}^2$ we deduce that $|m_2| \sim 5.3\times 10^{-2}$ eV and $|m_3| \sim 7.3\times 10^{-2}$ eV. Thus, in this model, the mass spectrum of active neutrinos is roughly degenerate, although $\Delta m^2_{21} \ll \Delta m^2_{32} \simeq \Delta m^2_{31}$

Making use of the estimate (28), we get from Eqs. (23), (24) and (25) the following estimations compatible with neutrino experimental data:  

\begin{eqnarray}
P(\nu_e \rightarrow \nu_e)_{\rm sol}\;\;\; & \sim & 1 - 0.83 \sin^2 (x _{21})_{\rm sol} - 0.069 \,,  \\
P(\bar{\nu}_e \rightarrow \bar{\nu}_e)_{\rm Chooz} & \sim & 1 - 0.069 
\end{eqnarray}

\ni and

\begin{equation} 
P(\nu_\mu \rightarrow \nu_\mu)_{\rm atm} \simeq 1 -  \sin^2 (x_{31})_{\rm atm} - 0.014 \,,
\end{equation}

\ni where $\theta_{12} \sim 33^\circ $. The neglected quadratic terms in $s^2_{14}, s^2_{25}$ and  $s^2_{36}$ would make the values of the shifts 0.069 and 0.014 as well as the oscillation amplitudes 0.83 and 1 in Eqs. (29), (30) and (31) a little bit smaller.

For larger LSND effect the parameter $s^2_{14}$ is larger, and thus the small deviations in Eqs. (29), (30) and (31) from pure three-active-neutrino oscillations grow, becoming more significant.

In conclusion, we have shown in this note on the ground of a simple $6\times 6$ neutrino mixing model that one of three conventional sterile (righthanded) neutrinos, if light enough, may be consistently used for explaining a small LSND effect. Then, it is still considerably heavier than the three active (lefthanded) neutrinos, so that a kind of a {\it soft} seesaw mechanism can work. 

The usual condition that the Majorana lefthanded mass matrix $M^{(L)}$ ought to vanish, {\it implies} the smallness of active-neutrino masses versus sterile-neutrino masses, when three mixing angles $\theta_{14}, \theta_{25}, \theta_{36}$ between both sorts of neutrinos are small (more precisely, $\theta_{14}, \theta_{25}, \theta_{36}$ are mixing angles between active neutrinos and the corresponding conventional sterile antineutrinos). In the present model, the mass spectrum of active neutrinos comes out roughly degenerate, lying in the range (5 --- 7.5) $\times 10^{-2}$ eV, {\it if} there is a {\it small} LSND effect with the amplitude of the order $10^{-3}$ and with the mass-squared splitting $\sim 1\;{\rm eV}^2$. 

\vfill\eject

~~~~

\vspace{0.5cm}

{\centerline{\bf References}}

\vspace{0.5cm}

{\everypar={\hangindent=0.7truecm}
\parindent=0pt\frenchspacing

{\everypar={\hangindent=0.7truecm}
\parindent=0pt\frenchspacing

~[1]~Q.R. Ahmad {\it et al.} (SNO Collaboration), {\it Phys. Rev. Lett.} {\bf 87}, 071301 (2001); {\tt nucl--ex/0309004}.

\vspace{0.2cm}

~[2]~Y. Fukuda {\it et al.} (SuperKamiokande Collaboration), {\it Phys. Rev. Lett.} {\bf 81}, 1562 (1998); {\it Phys. Lett.} {\bf B 467}, 185 (1999).

\vspace{0.2cm}

~[3]~M.H. Ahn {\it et al.} (K2K Collaboration), {\it Phys. Rev. Lett.} {\bf 90}, 041801 (2003).

\vspace{0.2cm}

~[4]~K. Eguchi {\it et al.} (KamLAND Collaboration), {\it Phys. Rev. Lett.} {\bf 90}, 021802 (2003).

\vspace{0.2cm}

~[5]~For a recent review {\it cf.} V. Barger, D. Marfatia, K. Whisnant, {\it Int. J. Mod. Phys.} {\bf E 12}, 569 (2003); M.~Maltoni, {\tt hep--ph/0401042}; and references therein.

\vspace{0.2cm}

~[6]~M. Apollonio {\it et al.} (Chooz Collaboration), {\it Eur. Phys. J.} {\bf C 27}, 331 (2003).

\vspace{0.2cm}

~[7]~C. Athanassopoulos {\it et al.} (LSND Collaboration), {\it Phys. Rev. Lett.} {\bf 77}, 3082 (1996); {\it Phys. Rev. } {\bf C 58}, 2489 (1998); A. Aguilar {\it et al.}, {\it Phys. Rev.} {\bf D 64}, 112007 (2001).

\vspace{0.2cm}

~[8]~H. Murayama, T. Yanagida, {\it Phys. Rev.} {\bf B 52}, 263 (2001); G. Borenboim, L.~Borissov, J. Lykken, A.Y. Smirnov, {\it J.High Energy Phys.} {\bf 0210}, 001 (2002); G. Borenboim, L. Borissov,  J. Lykken, {\tt hep--ph/0212116}. 

\vspace{0.2cm}

~[9]~A.O. Bazarko {\it et al.}, {\tt hep--ex/9906003}.

\vspace{0.2cm}

[10]~V. Barger, S. Pakvasa, T.J. Weiler, K. Whisnant, {\it Phys. Rev.} {\bf D 58}, 093016 (1998); {\it Phys. Lett.} {\bf B 437}, 107 (1998); M. Maltoni, T. Schwetz, M.A.~Tortola, J.W. Valle, {\it Nucl. Phys.} {\bf B 643}, 321 (2002); and references therein.

\vspace{0.2cm}

[10]~M. Sorel, J. Conrad, M. Shaevitz, {\tt hep--ph/0305255}.

\vspace{0.2cm}

[12]~W. Kr\'{o}likowski, {\tt hep--ph/0402183}. This reference implies that in Nature, beside three generations of leptons and quarks (SM (15+1)-plets), there are exactly two light sterile neutrinos (extra SM singlets); for a theoretical background {\it cf.} W. Kr\'{o}likowski, {\it Acta Phys. Pol.} {\bf B 30}, 227 (1999), Appendix; also {\it Acta Phys. Pol.} {\bf B 33}, 2559 (2002). In the present paper that is an alternative for Ref. [12], these two singlets are assumed to be completely decoupled from the Standard Model and so, invisible.

\vfill\eject

\end{document}